\documentclass{ws-procs975x65}            
\usepackage{float}%
\usepackage{amsmath}%
\begin{document}
\title{Recent results in CDT quantum gravity}

\author{Jan Ambjorn$^{a,b}$, Daniel Coumbe$^{a,c}$, Jakub Gizbert-Studnicki$^c$ and Jerzy Jurkiewicz$^c$}

\address{$^a$The Niels Bohr Institute, Copenhagen University,\\
Blegdamsvej 17, DK-2100 Copenhagen, Denmark.\\
E-mail: ambjorn@nbi.dk, Daniel.Coumbe@nbi.ku.dk}


\address{$^b$IMAPP, Radboud University, \\ Nijmegen, PO Box 9010, The Netherlands.}

\address{$^c$Faculty of Physics, Astronomy and Applied Computer Science,\\ 
Jagiellonian University, ul. prof. Stanislawa Lojasiewicza 11,\\ 
Krakow, PL 30-048, Poland\\
E-mail: jakub.gizbert-studnicki@uj.edu.pl, jerzy.jurkiewicz@uj.edu.pl}

\begin{abstract}

We review some recent results from the causal dynamical triangulation (CDT) approach to quantum gravity. We review recent observations of dimensional reduction at a number of previously undetermined points in the parameter space of CDT, and discuss their possible relevance to the asymptotic safety scenario. We also present an updated phase diagram of CDT, discussing properties of a newly discovered phase and its possible relation to a signature change of the metric.

\end{abstract}


\bodymatter

\section{Introduction}\label{intro}

Assuming little more than standard quantum field theory and general relativity, the causal dynamical triangulation (CDT) approach to quantum gravity has produced a number of significant results.~\cite{Ambjorn:2007jv,Ambjorn:2005db,Ambjorn:2011cg} The aim of this work is to highlight some more recent CDT results and discuss their potential importance in the future development of CDT and related fields.

CDT attempts to formulate a nonperturbative description of quantum gravity via a sum over all possible spacetime geometries. In CDT the geometry of spacetime is defined by an ensemble of locally flat $n$-dimensional simplices connected by their mutual $\left(n-2\right)$-dimensional faces. An important characteristic of CDT is the foliation of the lattice into space-like hypersurfaces of fixed topology, which enforces causality at a fundamental level. The path integral measure ensures the causality condition is respected by only including geometries that can be foliated in this way.    

The CDT action is

\begin{equation} 
S_{E}=-\left(\kappa_{0}+6\Delta\right)N_{0}+\kappa_{4}\left(N_{4}^{(4,1)}+N_{4}^{(3,2)}\right)+\Delta\left(2N_{4}^{(4,1)}+N_{4}^{(3,2)}\right),
\end{equation} 

\noindent where $\kappa_{0}$ is related to the inverse bare Newton's constant $G_{N}$, $\Delta$ is defined via the ratio of the length of space-like and time-like links on the lattice, and $\kappa_{4}$ is related to the cosmological constant.~\cite{Regge:1961px} $N_{4}^{(4,1)}$ is the number of 4-dimensional simplices of type $(4,1)$ and $N_{4}^{(3,2)}$ is the number of type $(3,2)$,  where the notation $(i,j)$ refers to the number of vertices $i$ on hypersurface $t$, and the number of vertices $j$ on hypersurface $t\pm1$. $\kappa_{4}$ is fixed so that an infinite volume limit can be taken. This leaves just two parameters, $\kappa_{0}$ and $\Delta$, which can be independently varied in order to explore the parameter space.

\section{New dimensional reduction results and asymptotic safety}

As is well known, gravity as a perturbative quantum field theory is not renormalizable by power counting.~\cite{Goroff:1985th} However, Weinberg suggested that gravity may be nonperturbatively renormalizable via the asymptotic safety scenario,~\cite{Weinberg79} which would require a finite number of couplings to terminate at a high-energy ultraviolet fixed point (UVFP). 

However, there exists an argument that may present a serious roadblock to realising the asymptotic safety scenario.~\cite{Banks:2010tj,Shomer:2007vq} The argument compares the high energy density of states expected for a theory of gravity with that of a conformal field theory. The entropy of a renormalizable theory is thought to scale according to 

\begin{equation}
S\sim E^{\frac{d-1}{d}}.
\label{eq:Entropy2}
\end{equation} 

\noindent However, the high energy spectrum of gravity is thought to scale according to

\begin{equation}
S\sim E^{\frac{d-2}{d-3}}.
\label{eq:Entropy3}
\end{equation}

\noindent Equation~(\ref{eq:Entropy2}) and~(\ref{eq:Entropy3}) disagree when $d=4$.~\cite{Banks:2010tj,Shomer:2007vq} This mismatch would then suggest that gravity cannot be formulated as a renormalizable quantum field theory in 4-dimensional spacetime.\footnote{However, the arguments leading to relations ~(\ref{eq:Entropy2}) and~(\ref{eq:Entropy3}) are highly controversial: specifically, the reasoning leading to Eq.~(\ref{eq:Entropy2}) has been criticized in Ref.~\refcite{Falls:2012nd}, and the assumptions leading to Eq.~(\ref{eq:Entropy3}) have been questioned in Refs.~\refcite{Percacci:2010af,Koch:2013owa}.} We now review recent spectral dimension results in CDT, and discuss how they may resolve this potential problem. 


The spectral dimension $D_{S}$ provides a measure of the effective fractal dimension of a manifold over varying length scales and is defined by

\begin{equation}
D_{S}=-2\frac{\rm{d}\rm{log}P_{r}}{\rm{d}\rm{log}\sigma},
\end{equation}

\noindent where $P_{r}$ is the probability that a random walk will return to the origin after $\sigma$ diffusion steps. Fitting the scale dependent spectral dimension data to the functional form $D_{S}=a-\frac{b}{c+\sigma}$ for the specific bare parameters $\kappa_{0}=2.2$ and $\Delta=0.6$ gives a long distance spectral dimension of $D_{S}\left(\sigma \rightarrow \infty \right)=4.02 \pm 0.10$ and a short distance spectral dimension of $D_{S}\left(\sigma \rightarrow 0 \right)=1.80 \pm 0.25$.~\cite{Ambjorn:2005db} This result led to the claim that the dimension of spacetime may dynamically reduce from 4-dimensions on large distances to 2-dimensions at the Planck scale, a result that has since been observed in numerous other approaches to quantum gravity.

\begin{figure}
  \centering
  \includegraphics[width=0.7\linewidth]{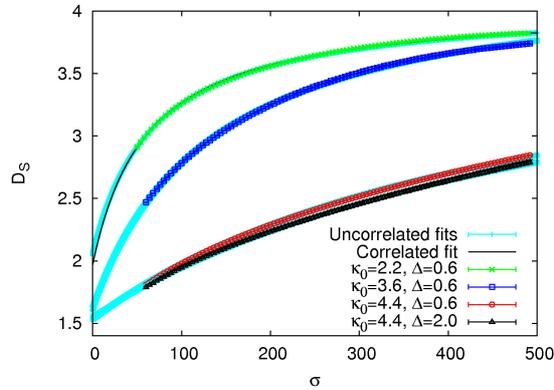}
  \caption{\small The spectral dimension $D_{S}$ as a function of the diffusion time $\sigma$ for different points in phase C of CDT.} 
\label{spec1}
\end{figure}

A more recent study determines the spectral dimension for a range of different bare parameters, $\kappa_{0}$ and $\Delta$, using a number of different lattice volumes and including a careful estimation of the associated statistical and systematic errors. This new study finds a short distance spectral dimension significantly more consistent with $D_{S}\left(\sigma \rightarrow 0 \right)=3/2$ than with $D_{S}\left(\sigma \rightarrow 0 \right) = 2$ for the majority of the sampled points within phase C of CDT, as shown in Fig.~\ref{spec1} and summarised in Table~\ref{BPTable}. This new study, however, does confirm the result $D_{S}\left(\sigma \rightarrow 0 \right) \simeq 2$ for the same bare parameters used in the previous study. 

\begin{table}
\tbl{\small A table of the long and short distance spectral dimension for different points in phase C of CDT.}
{\begin{tabular}{@{}cccccc@{}}\toprule
$(\kappa_{0},\Delta)$ & $N_{4,1}$ & $D_{S}(\infty)$ & $D_{S}(0)$ & $\sigma$ tension with $D_{S}(0)=2$ \\ \colrule
$\left(2.2,0.6\right)$ & 160,000 & $4.05\pm0.17$ & $1.970\pm0.266$ & 0.1 \\ \colrule
$\left(3.6,0.6\right)$ & 160,000 & $4.31\pm0.32$ & $1.576\pm0.093$ & 4.5 \\ \colrule
$\left(4.4,0.6\right)$ & 160,000 & $4.12\pm0.16$ & $1.534\pm0.058$ & 8.0 \\ \colrule
$\left(4.4,2.0\right)$ & 300,000 & $4.14\pm0.12$ & $1.540\pm0.060$ & 7.7 \\ \botrule
\end{tabular}}
\label{BPTable}
\end{table}


The fact that the small distance spectral dimension is consistent with $D_{S}\left(\sigma \rightarrow 0 \right)=3/2$ for the majority of points sampled in phase C of CDT may have important implications for the asymptotic safety scenario. We point out that Eqs.~(\ref{eq:Entropy2}) and~(\ref{eq:Entropy3}) are equal if $d=3/2$, which is precisely the value found for the short distance spectral dimension in CDT. Therefore, if the dimension of spacetime reduces to 3/2 in the high energy limit, as suggested by the CDT data, then the roadblock to asymptotic safety is removed, an idea first proposed within the context of Euclidean dynamical triangulations.~\cite{Laihobb}

\section{New phase diagram}

\begin{figure}
  \centering
  \includegraphics[width=0.6\linewidth]{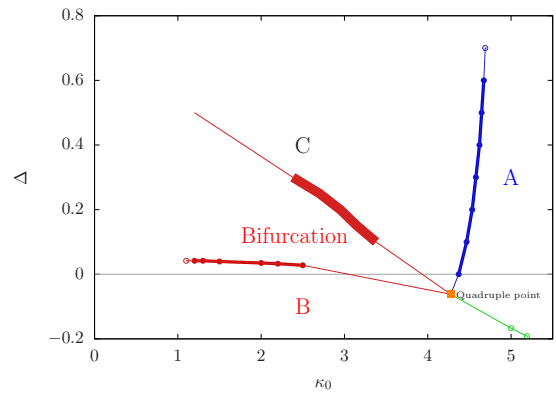}
  \caption{\small The updated phase diagram of 4-dimensional CDT.} 
\label{newpd}
\end{figure}

Until recently the phase diagram of CDT was thought to consist of only three phases A, B and C (see Fig.~\ref{newpd}). However, a new phase of CDT has recently been discovered, the so-called bifurcation phase.~\cite{Ambjorn:2014mra} The first evidence for the new phase came from analysing the effective transfer matrix.~\cite{Ambjorn:2014mra} The transition between phase C and the bifurcation phase has since been approximately located for a range of bare parameters, as shown in Fig.~\ref{newpd}.   

\begin{figure}[H]
  \centering
  \includegraphics[width=0.6\linewidth]{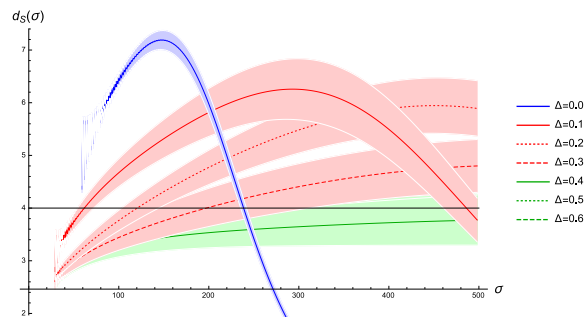}
  \caption{\small The spectral dimension as a function of diffusion time $\sigma$ for fixed $\kappa_{0}=2.2$ and various $\Delta$ values.} 
\label{spec2}
\end{figure}

The bifurcation phase appears to have a distinctly different geometry than phase C, with the spectral dimension becoming significantly larger than 4 for $\Delta$ values within the bifurcation phase, as indicated in Fig.~\ref{spec2}. 


\section{Signature change of the metric in CDT?}

The transfer matrix (TM) of CDT is the transition amplitude to go from spatial volume $n$ at discrete time $t$ to spatial volume $m$ at time $t+1$, while integrating out all other degrees of freedom. The TM has been shown to accurately reproduce the volume profile and quantum fluctuations in phase C.~\cite{Ambjorn:2014mra} Here, we apply the TM to analyse the properties of the newly discovered bifurcation phase.

The TM in phase C is accurately parametrised by

\begin{equation}
\langle n|M_C|m \rangle = \underbrace{\exp\Bigg[- \frac{1}{\Gamma} \frac{(n-m)^2}{ (n+m)}\Bigg] }_{\text{kinetic part}}  \underbrace{\exp\Bigg[ - \mu \left(\frac{n+m}{2}\right)^{1/3} + \lambda \left(\frac{n+m}{2}\right) \Bigg] }_{\text{potential part}} ,
\label{TMC}
\end{equation}
and in the bifurcation phase by
\begin{equation}
\langle n|M_B|m \rangle =\left[ \exp \left( -\frac{1}{\Gamma}\frac{\Big((n-m) - \big[c_0 (n+m - s_b) \big]_+\Big)^2}{n+m}\right) + \right.
\label{TMB}
\end{equation}
$$
+ \left. \exp \left( -\frac{1}{\Gamma}\frac{\Big((n-m) + \big[c_0 (n+m - s_b) \big]_+\Big)^2}{n+m}\right)\right] V[n+m] \ ,
$$

\noindent where $\Gamma$, $\mu$ and $\lambda$ are parameters related to Newton's constant, the size of the CDT universe and the cosmological constant, respectively. The bifurcation phase transition is associated with the parameters $s_b\to \infty$ and $c_0\to 0$, whereby Eq.~(\ref{TMB}) transforms into Eq.~(\ref{TMC}). Assuming that in the vicinity of the phase transition the potential term $V[n+m]$ in the bifurcation phase  approximates that of phase C, one can expand Eq. (\ref{TMB})   in powers of ${ 2 c_0}  (n-m)/ {\Gamma} \ll 1 $ to obtain the effective Lagrangian

\begin{equation}
L_B[n,m] \approx \frac{1}{\Gamma} \left(1-\frac{2 c_0^2(n+m)}{\Gamma}\right) \frac{(n-m)^2}{n+m} + \mu \left(\frac{n+m}{2}\right)^{1/3} - \left(\lambda  -\frac{c_0^2}{\Gamma}\right)   \left(\frac{n+m}{2}\right)
.
\label{lagrangian}
\end{equation}

For sufficiently large volumes $n+m > \Gamma /(2c_{0}^{2})$ we find that crossing the bifurcation transition (see Fig.~\ref{newpd}) makes the kinetic term in Eq.~(\ref{lagrangian}) flip sign. Therefore, it is tempting to interpret the transition between phase C and the bifurcation phase as a sign of an effective signature change from a Lorentzian metric in phase C to a Euclidean metric in the bifurcation phase. A similar result based on the spectral dimension of CDT has also been presented elsewhere.~\cite{Coumbe:2015zqa} Although this result is suggestive it is far from conclusive and further study is needed.   

\section{Conclusions}

We have reviewed three new results of CDT quantum gravity. Firstly, we have presented recent dimensional reduction results for a number of previously undetermined points in the parameter space of CDT. We find a large distance spectral dimension consistent with 4, as found previously,~\cite{Ambjorn:2005db} but a short distance spectral dimension that is more consistent with 3/2, rather than 2 as previously thought.~\cite{Ambjorn:2005db} We highlight the possible importance of this result for the asymptotic safety scenario, as first proposed in Ref.~\refcite{Laihobb} within a different context. Secondly, we present an updated picture of the phase diagram of CDT, including a description of the newly discovered bifurcation phase. Thirdly, we discuss possible evidence of a signature change of the metric due to the existence of the newly discovered bifurcation phase.

\section{Acknowledgments}

D.C and J.J wish to acknowledge the support of the grant DEC-2012/06/A/ST2/00389 from the National Science Centre Poland. J.G-S acknowledges the National Science Centre Poland (NCN) support via the grant DEC-2012/05/N/ST2/02698.


\small
\bibliographystyle{unsrt}
\bibliography{Master}


\end{document}